\documentstyle[preprint,eqsecnum,aps]{revtex}
\tightenlines
\input boxedeps.tex
\SetRokickiEPSFSpecial  
\HideDisplacementBoxes

\newcommand{\g}{\gamma}

\newcommand{\da}{\dagger}  
\newcommand{\be}{\begin{equation}}
\newcommand{\eq}{\end{equation}}
\newcommand{\Tr}{{\rm \, Tr \!}}    

\begin{document}
\title{Mesons on a  transverse
lattice (Revised 08/01)}

\author{Simon Dalley}
\address{Centre for Mathematical Sciences, Cambridge University \\
Wilberforce Road, Cambridge CB3 0WA, England}

\maketitle

\begin{abstract}
The meson eigenstates of the light-cone
Hamiltonian in a coarse transverse lattice gauge theory are investigated.
Building upon previous work in pure gauge theory, 
the Hamiltonian and its Fock space are expanded in powers of 
dynamical fields. In the leading approximation, the couplings
appearing in the Hamiltonian are renormalised 
by demanding restoration of space-time symmetries broken 
by the cut-off. 
Additional requirements from chiral symmetry are discussed and
difficulties in imposing them from first principles
in the leading approximation are noted. 
A phenomenological calculation is then performed, in which 
chiral symmetry in spontaneously broken form is 
modelled by imposing the physical $\pi$-$\rho$ mass splitting as a
constraint. The light-cone
wavefunctions of the resulting Hamiltonian are used
to compute decay constants, form factors and quark momentum and spin 
distributions for the pion and rho mesons.
Extensions beyond leading order, and the implications for first 
principles calculations, are briefly discussed.

\end{abstract}

\pacs{Valid PACS appear here.
{\tt$\backslash$\string pacs\{\}} should always be input,
even if empty.}
\narrowtext


\section{Introduction}
\label{intro}
Many questions of a dynamical nature in QCD cannot easily be
addressed in conventional lattice quantisation schemes \cite{wilson0,kogsus}.
Light-cone quantisation is a Hamiltonian framework suitable for
highly-relativistic bound and scattering states \cite{rev,wilson1}, 
and would be ideal
for extending the kinds of problem that may be tackled. The
corresponding light-cone wavefunctions are frame-invariant, contain no
disconnected vacuum contributions, and 
matrix elements of suitable operators can be related to virtually any
hadronic process. 
In this work, the light-cone 
Hamiltonian for a version of lattice QCD with quarks is investigated. 
It builds upon previous work of the author with B. van de Sande on the
pure gauge theory \cite{dv0,dv1,dv2},
following an original idea of Bardeen {\em et. al.} \cite{bard1,bard2}. 
The particular lattice gauge theory and its approximation  considered
here was first proposed in ref.\cite{mat0}.

All the transverse  lattice treatments of non-abelian 
gauge theory cited above are
formulated in terms of disordered link-fields $M$, which transform
gauge covariantly but are not themselves
restricted to the gauge group. 
As such, they are  expected to be well-suited to  
lattice spacings $a$ that are coarse-enough (compared to $\Lambda_{\rm
QCD}$) for the use of  disordered elementary degrees of freedom.
The question then arises as to how to determine the effective Hamiltonian.
The expected level of coarseness probably 
rules out any reliable use of perturbation theory on the basis of
asymptotic freedom. The Hamiltonian must be renormalised with
a non-perturbative technique. 
Pure QCD may be defined by gauge and Poincar\'e symmetries, together with
an explicit connection to the asymptotically free
continuum limit. 
In transverse lattice gauge theory, one discretizes two (transverse) space
directions on a lattice, say ${\bf x} = \{x^1, x^2 \}$, while
maintaining lattice gauge invariance. 
In refs.\cite{dv1,dv2} it was suggested that, since the other space-time
directions $\{x^0, x^3 \}$ remain continuous, tuning the Hamiltonian to
restore Poincar\'e symmetry   at finite transverse
lattice spacing would be enough to study spectral quantities
in QCD. An approximation scheme for implementing this idea
is the colour-dielectric expansion. For sufficiently massive fields
$M$, low-lying hadronic light-cone  wavefunctions  may converge
rapidly in parton number, making expansion of the Hamiltonian
in powers of $M$ viable.
This method was tested and symmetry requirements alone 
produced a well-defined trajectory of couplings to 
leading order of the power expansion of the Hamiltonian.
Glueball spectra on this trajectory scaled and agreed with results
from other methods (the most  accurate agreement has been  found
in $2+1$-dimensional studies \cite{dv3}).

A similar method is pursued here in the light-meson sector, where
a  number of new features arise. 
In anticipation of a constituent structure appearing
for boundstates when the ultra-violet cut-off is lowered
sufficiently, expansion of the Hamiltonian
in powers of dynamical fields, both disordered links and quarks,
will again be the basis of the approximation scheme. 
In this work, the expansion is also extended to the Hilbert space itself;
an additional Tamm-Dancoff cut-off will be put
on the number of partons in a Fock space. The convergence
of these approximations will not be studied here; rather, as a
necessary first step, the lowest
order of approximation will be solved in order to see if a reasonable
phenomenology results at this level. 
The cut-off Hamiltonian will be renormalised non-perturbatively
by searching the space of couplings
for a region where spacetime  symmetries violated by
the cut-off are being restored.
Unlike in pure gauge theory, however, one must 
also confront approximate chiral
symmetry (i.e. one must fix the quark mass).
Practical problems when trying to impose both Poincar\'e and
chiral symmetry in the
leading approximation are noted, which may require higher 
orders and/or renormalisation of operators other than the 
Hamiltonian to overcome. For this work,
chiral symmetry in spontaneously broken form is therefore modelled by 
demanding  the $\pi$-$\rho$ mass  splitting as
a phenomenological constraint.
With this caveat, 
a well-defined region in coupling space that enhances covariance of
light-meson boundstates is found.

Using the bare expressions for currents (lowest order in an
expansion in dynamical fields),
the light-cone eigenfunctions of the  Hamiltonian renormalised
and constrained in this way
are used to compute decay constants, form factors and quark momentum and spin
distributions of the pion and rho.
These are compared with 
experimental results, where available, and other 
phenomenological  and  lattice results. 
The leading approximation is found to be a
useful quantitative starting point for the calculation of such observables.
Issues in going beyond the leading approximation and how this
may allow one to use symmetry rigorously as a renormalisation method, 
are briefly
discussed.

\section{Transverse lattice}

\subsection{Overview}

Light-cone quantisation is a Hamiltonian method that treats $x^+= (x^0
+ x^3)/\sqrt{2}$ as
canonical time and $\{x^-=(x^0-x^3)/\sqrt{2},
{\bf x}\}$ as the spatial variables.
Lorentz indices $\mu, \nu $ are
split into light-cone indices $\alpha,\beta \in \{+,-\}$
and transverse indices $r,s\in \{1,2\}$. The metric
has non-zero entries $g^{+-} = - g^{11} = -g^{22} = 1$. 
In transverse lattice gauge theory, link variables
$M_r(x^+,x^-,{\bf x})$  are associated
with the link from ${\bf x}$ to ${\bf x} + a \hat{\bf r}$ at fixed
$x^+$
and $x^-$. They  physically represent the (averaged) flux
between these points. Continuum $SU(N)$ gauge potentials 
$A_{\alpha}(x^+,x^-,{\bf x})$ and fermions fields $\Psi(x^+,x^-,{\bf x})$ 
are associated to a transverse plane ${\bf x} = {\rm const.}$. A slice through
spacetime at fixed $x^+$ is illustrated in figure 1. 
The field variables transform under transverse lattice gauge
transformations $V(x^+,x^-,{\bf x}) \in SU(N)$ associated with the
transverse planes as
\begin{eqnarray}
        A_{\alpha}({\bf x}) & \to & V({\bf x}) A_{\alpha}({\bf x}) 
        V^{\da}({\bf x}) + {\rm i} \left(\partial_{\alpha} V({\bf x})\right) 
        V^{\da}({\bf x}) \ ,  \nonumber \\
        M_r({\bf x}) &  \to & V({\bf x}) M_r({\bf x})  
        V^{\da}({\bf x} + a\hat{\bf r})  \label{sym} \ , \\
        \Psi({\bf x}) & \to & V({\bf x})\Psi({\bf x}) \ . \nonumber  
\end{eqnarray}
The most general transverse lattice 
Lagrangian $L$ will consist of all operators invariant
under lattice gauge symmetries (\ref{sym}), Poincar\'e symmetries
manifestly
preserved by the lattice cut-off, and renormalisable by dimensional
counting with respect to the continuum co-ordinates $x^{\alpha}$.
It is important to note that $M$ need not be restricted to $SU(N)$
 \cite{bard1}. 
It can be taken as a general $N$x$N$ complex matrix, provided it still
gauge transforms covariantly.

For  QCD, in the transverse continuum limit $a \to 0$ one expects
the couplings in $L$ to run with $a$ such that 
$M$ is forced into $SU(N)$ \cite{wilson0}. 
Writing $M=HU$, where $H$ is an hermitian matrix,
as the transverse continuum limit is approached, 
$H$ should be expanded about a 
vacuum value $H = H_0 + \tilde{H}$. The fluctuation $\tilde{H}$
should be made infinitely heavy  by $L$ as it runs to $a =0$. 
Unfortunately,  
the light-cone quantisation of non-linear degrees of freedom $U$ 
and vacuum modes $H_0$ is  very 
awkard. Even if one formulates the theory entirely in terms of only
unitary variables $U$ from the beginning, the light-cone quantisation
is still highly non-trivial \cite{griff}.
However, in the region of coupling space
where $H_0 = 0$, $M$ behaves as a massive linear degree of freedom and
light-cone quantisation is straightforward.
The colour-dielectric hypothesis suggests that the renormalised 
trajectory \cite{wilson2}, on which sufficiently long-distance
observables take the same values as in the continuum limit,
passes into this region at sufficiently large $a$ \cite{dv0,mack}.
Since weak-coupling perturbation theory is unlikely to be reliable
for the construction of this trajectory at large $a$,
it was suggested to use symmetry to reconstruct it \cite{dv2,dv3}. 
In the case of a
transverse lattice regulator, it is plausible that with the
most general cut-off $L$, gauge and Poincar\'e symmetry are sufficient
to identify a unique trajectory associated to QCD. 
Whether or not this is correct, it nevertheless motivates an
approximate treatment of this idea.

In reality, one must
truncate the number of operators appearing in the most general $L$ and also
introduce further
cut-offs on the Hilbert space, beyond the transverse lattice cut-off $a$. 
The subset of symmetry criteria one chooses are now obviously 
influenced by the approximations made.
The truncation of $L$, in the presence of 
only the cut-off $a$, will be based upon
the relative importance of
different field configurations expected in low-lying hadronic boundstates.
This substitutes for the fact that one cannot make use of 
dimensional counting with 
respect to transverse co-ordinates ${\bf x}$, since 
$a$ is not assumed to be small.
It then makes sense to test symmetries only in the lowest-energy 
observables. Eventually, this procedure should be repeated,
adding more operators to $L$ and testing symmetries at ever
higher energies.
It is possible that the approximation scheme employed, although
systematic and physically motivated, is poor. That is, it restricts
one to couplings to operators that poorly approximate a renormalised
trajectory. This can only be decided by explicitly checking for
restoration of symmetries violated by the cut-off.
The approximations will be further modified by the 
presence of additional cut-offs on the Hilbert space.
The simplest  non-trivial approximation
to the QCD Lagrangian and Hilbert space is used as the basis for an  
explicit computation in this paper.
Issues in going beyond this approximation are also briefly discussed.

In the rest of this section, approximation of the Hamiltonian (II.B),
Hilbert space (II.C), and chiral symmetry issues (II.D) are
addressed. The calculation within the leading approximation is
detailed in section III, first with the introduction of
explicit symmetry criteria (III.A),  then applications to
phenomenology (III.B \& C).

\subsection{Approximating the Hamiltonian}
\label{hamilton}
Starting from the most general transverse lattice Lagrangian $L$,
we wish to arrive at a systematically improvable approximation to 
the light-cone Hamiltonian $P^-$. First, recall the
(continuum) proper Poincar\'e generators in light-cone co-ordinates 
\cite{kogsop} for
a set of single-component fields labelled generically $\phi_l$.
At a particular light-cone time $x^+$,
\begin{eqnarray}
P^{\nu} & = & \int dx^- d^2{\bf x} \ T^{+ \nu} \\
M^{\mu \nu} & = & \int dx^- d^2{\bf x} \left(x^{\mu} T^{+\nu}
-x^{\nu}T^{+\mu} + \pi_l \Sigma_{lm}^{\mu \nu} \phi_m \right) \ ,
\end{eqnarray}
where $T$ is the energy-momentum 
\be
T^{+\mu} = \pi_l \partial^{\mu} \phi_l - g^{+ \mu} L \ ,
\eq
$\pi_l$ are the canonical momenta
\be
\pi_l = {\delta L \over \delta (\partial_{+} \phi_l)} \ ,
\eq
and the spin part for spinor or vector fields respectively is
\be
\Sigma_{lm}^{\mu \nu} = {1 \over 4} [\gamma^{\mu}, \gamma^{\nu}]_{lm}
\ \ \ {\rm or } \ \ \ g^{\mu}_{l}g^{\nu}_{m}-g^{\mu}_{m}g^{\nu}_{l} \
. 
\eq
The mass shell condition is 
\be
2 P^+ P^- = {\cal M}^2 + (P^1)^2 + (P^2)^2 \ .
\eq
There are also the discrete symmetries $\{P:  x^- \leftrightarrow x^+, {\bf x}
\to - {\bf x} \}$ and $\{ T: x^- \leftrightarrow -x^+ \}$, related to improper
Lorentz transformations. For the moment, no chiral symmetry is assumed.

The (classical) 
space-time symmetries will be classified into violated or preserved, and
kinematic or dynamic, in the presence of the transverse lattice 
cut-off $a$.
The most general Lagrangian may only contain lattice gauge-invariant
operators invariant under
preserved Poincar\'e symmetries. In the continuum,
the subset $\{ P^r, P^+, M^{+r}, M^{12},
M^{-+}\}$ can usually be chosen {\em kinematic}. 
That is, they can be chosen --- at a particular $x^+$ and
in a particular gauge perhaps --- to be 
independent of interactions, quadratic in independent dynamical fields. 
As many generators as possible will be maintained in kinematic form,
consistent with lattice gauge invariance. 
The imposition of a lattice cut-off on transverse directions means that
the continuous symmetries associated to 
$\{ P^r , M^{12}, M^{+r}\}$ are {\em violated} 
(although there is a discrete subset
of finite symmetry transformations that are preserved).
Demanding that the generators $\{P^+, M^{-+}\}$ of {\em preserved} symmetries
can be chosen kinematic, they can easily be used to define states of
non-zero longitudinal momentum. 
Although the symmetry generated by $M^{+r}$ is in general violated,
at a particular time $x^+=0$ it is recovered and kinematic, and 
will be used to define transversely boosted states at this time.
Of the remaining {\em dynamic}  generators $\{P^- , M^{-r}\}$, 
the symmetry generated by the Hamiltonian
$P^-$ is preserved, while
that generated by the boost-rotations $M^{-r}$ is violated.
Parity $P$ and time reversal $T$ are dynamic and preserved, though $PT$, 
hence charge congugation $C$ (or $G$-parity), is kinematic.

Demanding a kinematic $P^+$ restricts the possible kinetic
term for $L$ to 
\begin{eqnarray}
L_{\rm kin}  & = &  \sum_{{\bf x}} \int dx^- \sum_{\alpha, \beta = +,-}
\sum_{r=1,2} 
-{1 \over 2 G^2} \Tr \left\{ F^{\alpha \beta} F_{\alpha \beta} \right\}
 + {\rm i} \overline{\Psi} 
\g^{\alpha} (\partial_{\alpha} + {\rm i} A_{\alpha}) \Psi
\nonumber
\\
&& + \Tr\left\{[\left(\partial_{\alpha} +{\rm i} A_{\alpha} ({\bf x})\right)
        M_r({\bf x})-  {\rm i} M_r({\bf x})   A_{\alpha}({{\bf x}+a
 \hat{\bf r}})][{\rm h.c.}]\right\}  \nonumber
\end{eqnarray}
in which ${\rm h.c.}$ is hermitian conjugate and
light-cone gauge $A_{-} = 0$ is chosen. 
This partial
gauge fixing will render $A_{+}$ 
non-dynamical, independent of $x^{+}$ according to equations
of motion; half the components of the fermion spinor,
$\Psi^{-}=\gamma^{0} \gamma^{-} \Psi$, will also be non-dynamical (see below).

The potential term for $L$  will be expanded in 
powers of the dynamical fields $M$, $\Psi^{+}= \gamma^{0} \gamma^{+}
\Psi$ (after elimination of non-dynamical fields $A_{\pm}$
and $\Psi^{-}$).
Such a power expansion can only be
justified in a  region of coupling
space where these fields are sufficiently heavy that light-cone
wavefunctions
of interest converge quickly in parton number.
For sufficiently heavy fields, 
the lowest-mass lattice hadrons will  consist of a few partons 
$M$ and  $\Psi_{+}$
with little mixing into configurations with many partons. 
The physical motivation for expecting 
a renormalised trajectory to exist in the region of link-fields $M$
with positive mass squared is the 
the colour-dielectric picture of confinement
\cite{bard1,mack}. 
The dynamical generation of a  constituent quark mass in
QCD makes this scenario  physically plausible for quarks also.
In both cases, the ultra-violet cut-off on the
renormalised trajectory must be lowered sufficently for these
constituent approximations to be viable.
Dimensional counting in the continuous directions $(x^+,x^-)$,
together with the power expansion in dynamical fields, limits the
number of allowed operators to some extent. However, locality in $x^-$ 
cannot be assumed. Elimination of non-dynamical degrees of
freedom will inevitably introduce non-local terms. This could,
in principle, lead to $x^-$-dependent couplings \cite{wilson1}. Demanding
that
$P^-$ is derived from a gauge-invariant Lagrangian $L$ and that  
$P^+$ be kinematic appears to eliminate this
possibility however, at least at level of the lattice cut-off theory.

The simplest
approximation to the Lagrangian in QCD is then  
\begin{eqnarray}
L & = &  \sum_{{\bf x}} \int dx^- \sum_{\alpha, \beta = +,-}
\sum_{r=1,2} 
-{1 \over 2 G^2} \Tr \left\{ F^{\alpha \beta} F_{\alpha \beta} \right\}
 \nonumber
\\
&& + \Tr\left\{[\left(\partial_{\alpha} + {\rm i} A_{\alpha} ({\bf x})\right)
        M_r({\bf x})-  {\rm i} M_r({\bf x})   A_{\alpha}({{\bf x}+a
 \hat{\bf r}})][{\rm h.c.}]\right\}
\nonumber \\
&& - \mu_{b}^2  \Tr\left\{M_r M_r^{\da}\right\}
 + {\rm i} \overline{\Psi} 
\g^{\alpha} (\partial_{\alpha} + {\rm i} A_{\alpha}) \Psi - \mu_f
\overline{\Psi}\Psi 
\nonumber\\
&& +  {\rm i} \kappa_a \left( \overline{\Psi}({\bf x}) \g^{r} M_r({\bf x})
 \Psi({\bf x} + a \hat{\bf r}) 
- \overline{\Psi}({\bf x}) \g^{r} M_{r}^{\da}({\bf x}- a \hat{\bf r}) 
\Psi({\bf x} - a \hat{\bf r})
\right)\nonumber\\
&&
+ \kappa_s \left( \overline{\Psi}({\bf x}) M_r({\bf x})
 \Psi({\bf x} + a \hat{\bf r})+\overline{\Psi}({\bf x}) M_{r}^{\da}({\bf
x}- 
a \hat{\bf r})
 \Psi({\bf x} - a \hat{\bf r})\right) \label{ferlag}
\end{eqnarray}
where $G,\mu_f,\mu_b,\kappa_a,\kappa_s$ are coupling constants
and the $A_- = 0 $ gauge is taken.
This will be the starting point for  explicit calculations in this paper.
Detailed investigation of the corrections to this $L$, involving
higher powers of fields,
is left to future work.

In the chiral representation (Appendix A-2 of ref.\cite{IZ}),  
$\Psi^{\dagger} = (u_{+}^{*}, v_{+}^{*}, v_{-}^{*},
u_{-}^{*})/2^{1/4}$ decomposes
into left (right) movers $v$ ($u$) with a helicity
subscript $h = \pm$.
In light-cone gauge $A_- = 0$,
$A_{+}$ and $v_{\pm}$ satisfy constraint equations of motion 
which are used to eliminate them at the classical level. Defining 
\begin{eqnarray}
F_{h}({\bf x}) & = & - u_{h}({\bf x}) + {\kappa_s \over \mu_f}
\sum_{r} \left( M_r({\bf x}) u_{h}({\bf x}+ a \hat{\bf r}) + 
M_{r}^{\da}({\bf x}-a \hat{\bf r} )u_{h}({\bf x}- a \hat{\bf r})
  \right) \nonumber \\ 
&&+  
{h {\rm i} \kappa_a \over  \mu_f} \left\{ M_{1}({\bf x} ) u_{-h}({\bf
x}+ a \hat{\bf 1})
- h{\rm i}M_{2}({\bf x} )) u_{-h}({\bf x}+ a \hat{\bf 2}) \right. \nonumber \\
&&   \left. - M_{1}^{\da}({\bf x}- a \hat{\bf 1} ) u_{-h}({\bf x}- a 
\hat{\bf 1})
+ h{\rm i}M_{2}^{\da}({\bf x}- a \hat{\bf 2} ) u_{-h}({\bf x}- a \hat{\bf 2})
\right\}  \\
J^{+}({\bf x}) &=& {\rm  i} \sum_{r}
\left(
M_r ({\bf x}) \stackrel{\leftrightarrow}{\partial}_{-} 
M_r^{\da}({\bf x})  + M_r^{\da}({\bf x} - a\hat{\bf r}) 
\stackrel{\leftrightarrow}{\partial}_{-} M_r({\bf x} - a\hat{\bf r})
\right)  \nonumber \\
&& + \sum_{h} u_{h}({\bf x})u_{h}^{\dagger}({\bf x}) \ ,
\end{eqnarray}
the constraints are
\begin{eqnarray}
(\partial_{-})^2 A_{+} &  = & {G^2 \over 2} \left( J^+ - {1 \over N} \Tr
\ J^+ \right)\\
{\rm i}
\partial_{-} v_{h} & = & {{\mu_f} \over \sqrt{2} } F_{-h} \ \label{cons}.
\end{eqnarray}
Introducing gauge indices $\{ i,j \in \{1,2, \cdots N \}\}$, the
canonical momenta are $\pi_{i}(u_{h}) = {\rm i} u_{h,i}^{*}$,
$\pi_{ij}(M_r) = \partial_{-}M_{r,ij}^{*}$. Hence,
the generators of preserved symmetries at $x^+ = 0$ are
\begin{eqnarray}
 P^-  &  = &  \int dx^- \sum_{{\bf x}} 
   {G^2 \over 4} \left(\Tr\left\{ 
              J^{+} \frac{1}{({\rm i} \partial_{-})^{2}} J^{+} \right\}
            -{1 \over N} 
        \Tr\left\{ J^+  \right\} {1 \over ({\rm i}\partial_{-})^{2} }
     \Tr\left\{ J^+ \right\} \right)  \nonumber \\
&& + {\mu_{f}^{2} \over 2} \sum_{h} 
\left( F_{h}^{\da} {1 \over {\rm i} \partial_-}
F_h \right) 
 + \mu_{b}^2  \sum_{r=1}^{2}  \Tr\left\{M_r M_r^{\da}\right\} 
\ , \label{lcham}\\
P^+ & = &  \int dx^- \sum_{{\bf x}, s, h} 2 \Tr  
                        \left\{ \partial_- M_s({\bf x})  
                \partial_- M_s({\bf x})^{\da} \right\}
+ {\rm i} u_{h}^{*} \partial_{-} u_{h}
 \ , \label{mom} \\
M^{-+} & = &   \int dx^- \sum_{{\bf x}, s, h}  x^- \left\{ 
             2 \Tr\left\{ \partial_{-} M_s({\bf x}) 
             \partial_{-} M_s({\bf x})^{\da} \right\} + {{\rm i}\over 2}  
u_{h}^{*} \stackrel{\leftrightarrow}{\partial}_{-} u_{h} \right\}  \ ,
   \\
M^{+r} & = & -  \int dx^- \sum_{{\bf x}, s, h} 
      2 \left( x^r + \frac{a}{2} \delta^{rs}\right) 
                      \Tr  \left\{ \partial_- M_s({\bf x})  
                \partial_- M_s({\bf x})^{\da} \right\} +   {\rm i} x^r 
u_{h}^{*} \partial_{-} u_{h} \ .
\end{eqnarray}
The Hamiltonian (\ref{lcham}) contains cubic and quartic interactions 
in dynamical fields. 
At quartic order there would be
further terms, in addition to those in (\ref{lcham}), generated
from quartic terms allowed in the Lagrangian, that have been
neglected in (\ref{ferlag}). 
The lowest order of the  power expansion
would retain only the cubic interactions in $P^-$, but this approximation 
would not produce confinement. 
One is free to include some or all of the possible terms at quartic
order in $P^-$. As a minimal prescription,  only
the quartic terms arising from the $J^+ J^+$ interaction will be
retained, as they are be responsible for confinement \cite{bard1}. 
The effect of other quartic
terms is left for future investigation; in particular, it is likely that they
will be important for parity and chiral symmetry restoration, as 
discussed later.

\subsection{Cut-offs in Fock space}
\label{Fock}

In the quantum theory, there are (anti-)commutation
relations at fixed $x^+$
\begin{eqnarray}
        \left[M_{r,ij}(x^-,{\bf x}), 
        \partial_- M_{s,kl}^{*}(y^-,{\bf y})\right]
       &  = & {{\rm i} \over 2} \delta_{ik}\,\delta_{jl}\, \delta (x^- -y^-)
        \, \delta_{\bf x y} \,\delta_{rs} \ , \\
\left\{u_{h,i}(x^-,{\bf x}), 
       {\rm  i} u_{h',j}^{*}(y^-,{\bf y})\right\}
        & = & {\rm i}  \delta_{ij}\,  \delta (x^- -y^-)
        \,{\bf \delta_{x y}} \,\delta_{hh'} \ .
\end{eqnarray}
A convenient Fock space representation  employs longitudinal
momentum space but transverse position space
\begin{eqnarray}
 M_r(x^+=0,x^-,{\bf x})   &=&  
        \frac{1}{\sqrt{4 \pi }} \int_{0}^{\infty} {dk^+ \over \sqrt{ k^+}}
        \left( a_{-r}(k^+,{\bf x})\, e^{ -{\rm i} k^+ x^-}  +  
        a^{\da}_r(k^+,{\bf x})\, e^{ {\rm i} k^+ x^-} \right)   \; ,
        \label{expand}
\\
   \left[a_{\lambda,ij}(k^+,{\bf x}), 
        a_{\rho,kl}^{*}(\tilde{k}^+, {\bf y}\right] 
        & = & \delta_{ik}\, \delta_{jl}\, \delta_{\lambda \rho}\, 
        {\bf \delta_{x y}}\,\delta(k^+-\tilde{k}^+) \;, \\
   \left[a_{\lambda,ij}(k^+,{\bf x}),
        a_{\rho,kl}(\tilde{k}^+, {\bf y})\right] & = & 0 \;.
\end{eqnarray}
\begin{eqnarray}
 u_h(x^+=0,x^-,{\bf x})   &=&  
        \frac{1}{\sqrt{2 \pi }} \int_{0}^{\infty} dk^+ 
        \left( b_{h}(k^+,{\bf x})\, e^{ -{\rm i} k^+ x^-}  +  
        d^{*}_{-h}(k^+,{\bf x})\, e^{ {\rm i} k^+ x^-} \right)   \; ,\\
\left\{b_{h,i}(k^+,{\bf x}), 
        b_{h',j}^{*}(\tilde{k}^+, {\bf y}\right\} 
        & = & \delta_{ij}\,  \delta_{hh'}\, 
        {\bf \delta_{x y}}\,\delta(k^+-\tilde{k}^+) \;, \\
   \left\{b_{h,i}(k^+,{\bf x}),
        b_{h',j}(\tilde{k}^+, {\bf y})\right\} & = & 0 \;.
\end{eqnarray}
Here, $\lambda$ and $\rho \in \{ \pm 1, \pm 2\}$, $a_{\lambda,ij}^{*} = 
a_{\lambda,ji}^{\dagger}$, and similar anti-commutators exist for $d$.
The Fock space operator $a_{ r,ij}^{\dagger}(k^+,{\bf x})$ 
creates a link-parton with longitudinal momentum
$k^+$, carrying colour $i$ at ${\bf x}$ to $j$ at 
 ${\bf x} + a\widehat{\bf r}$. 
$b_{h,i}^{*}(k^+,{\bf x})$ creates a quark of helicity $h$, colour $i$,
momentum $k^+$ at site ${\bf x}$, while $d^*$ does the same for
anti-quarks.
This Fock space is already diagonal in $P^+$ and serves as a basis
for finding the eigenvalues of $P^-$.

The $1/\partial_{-}$ non-local terms in $P^-$ (\ref{lcham})
cause divergences
at small $k^+$.
However, the well-known elimination of such divergences by a suitable 
normal-ordering prescription in $1+1$-dimensional  gauge and Yukawa
theories \cite{dlcq}
carries over to the transverse lattice Hamiltonian at finite
$a$.
While the currents $J^+$ and $F$ are normal-ordered, the
Hamiltonian $P^-$ itself is not. This generates infinite self-inertias for the
$M$ and $\Psi^+$ partons, which are required for finiteness of the
eigenvalues of $P^-$.
Moreover, as can be seen from (\ref{lcham}),
the finite-energy physical states $|\psi\rangle$ satisfy
\be
\int dx^-  J^+ |\psi\rangle = \int dx^- F_{\pm} |\psi\rangle = \int dx^- 
F_{\pm}^{\dagger} |\psi\rangle = 0 \ . \label{phys}
\eq
The $J^+$ condition singles out Fock states invariant under residual
$x^-$-independent gauge transformations $V({\bf x})$ in the light-cone gauge
\cite{bard1}.  For these singlet configurations, there is linear
confinement of heavy sources for generic coupling choices in the
region of sufficiently large $\mu_{b}^{2}$ \cite{mat1}.
The $F$ conditions will relate Fock space wavefunction components 
at small $k^+$ momentum
with different numbers of partons, and are
the source of rising (Regge) behaviour of structure functions at 
small Bjorken $x$ \cite{ladder}.

In order to render the Fock space finite-dimensional, suitable for
study on a computer, it is
necessary to impose further cut-offs. DLCQ \cite{dlcq,old} will be
used to
discretize longitudinal momentum, by
compactifying  $x^-$ on 
circle of circumference ${\cal L} = 2 \pi K/ P^+$, where $K$
is a positive  integer, with
(anti)-periodic boundary conditions for $M$ ($u$). All $k^+ =
0$ modes will be dropped, whether or not the boundary conditions allow
it
.\footnote{This 
justifies {\em post hoc} the fact that boundary terms were not
discussed earlier.}
The DLCQ cut-off is applied directly to the Poincar\'e 
generators and 
results extrapolated to $K = \infty$.
However, since the zero modes are not picked up explicitly by the
$K \to \infty$ extrapolation, further finite renormalisation due to 
their absence is
expected. 
General arguments about the
kinds of finite renormalisation expected upon omission
of zero modes in light-cone QCD have been given in ref.\cite{wilson1}.
Previously preserved dynamical
symmetries could in general be violated by
DLCQ without zero modes, even in the limit $K \to \infty$. In
particular, parity invariance
is violated  by zero mode omission and some work
has been done on the corresponding renormalisation required 
in the context of simpler field
theories \cite{mat2}. This is one place where quark `kinetic' mass terms
can persist, even in a chiral theory, leading to a viable constituent
approximation.
The quartic terms neglected in $P^-$ (\ref{lcham}) are
needed to address parity restoration in general.

There are also zero modes associated to the transverse and
longitudinal gauge fields. To justify the power expansion, 
the link-field  mass $\mu_{b}^2$ was taken sufficently large. This ensures that
the $k^+ = 0$ mode of the (radial part  of the)
transverse fields $M$ is pushed to infinite energy, so  one
may indeed expand about $M=0$ (\ref{expand}).
On the other hand, the omission of zero modes of $A^{\alpha}$
could lead to additional finite renormalisations for
zero momentum transfer 4-parton interactions induced by the
instantaneous gluon exchange kernel $1/\partial_{-}^{2}$.
While the zero mode of $A_+$ can be gauged away at a
particular $x^+$,
the zero mode of $A_{-}$ cannot be gauge fixed to 
zero in the presence of periodic $x^-$ boundary conditions \cite{fix}.
There has been some study of the effects of explicitly retaining such a mode
in simpler field theories \cite{chris}, which indicate that there  may be
a small effect in the instantaneous gluon exchange   between
$M$ and $\Psi^+$. Parity symmetry again would be the obvious place
to look for fixing these zero-mode induced effects. 
No attempt
to incorporate further zero-mode induced operators into $P^-$ will
be made here.

Although DLCQ at fixed $K$ and without zero modes automatically  cuts off
the maximum number of partons in Fock space, it will be
convenient to impose a separate Tamm-Dancoff cut-off 
on this number. Ideally, this cut-off would be extrapolated,
along with the DLCQ one, as has been done in transverse lattice
studies of pure gauge theory \cite{dv2}. In the present work,
the Tamm-Dancoff cut-off will be {\em fixed} at the smallest
non-trivial value throughout
the calculation for simplicity. In the `one-link' truncation of Fock
space,  a general  ${\bf P} ={\bf 0}$ normalised
state can be expressed in
terms of orthonormal Fock states as
\begin{eqnarray}
|\psi(P^+)\rangle &= &  
 {1 \over \sqrt{\rm Vol}} \sum_{\bf x} a^2 \sum_{h,h'}
\int_{0}^{1} dx_1 dx_2 \ \delta (x_1 + x_2 -1) \nonumber \\  
&& \times \left\{ \psi_{hh'}(x_1,x_2) N^{-1/2} 
b_{h}^{\dagger}(x_{1}, {\bf x})d_{h'}^{*}(x_{2}, {\bf x})
|0\rangle \right\}\nonumber \\
&& + 
 {1 \over \sqrt{\rm Vol}} \sum_{\bf x} a^2 \sum_{h,h',r}
\int_{0}^{1} dx_1 dx_2 dx_3 \ \delta (x_1 + x_2 + x_3 -1) 
\\ 
&& \times \left\{ \psi_{h(r)h'}(x_1,x_2,x_3) N^{-1}
  b_{h}^{\dagger}(x_{1}, {\bf x})a^{\dagger}_{r}(x_{2},{\bf x})
d_{h'}^{*}(x_{3}, {\bf x}+ a\hat{\bf r}) |0\rangle \right. \nonumber \\
&& \left. +  \psi_{h(-r)h'}(x_1,x_2,x_3)  
N^{-1} b_{h}^{\dagger}(x_{1}, {\bf x}+ 
a\hat{\bf r})
a^{\dagger}_{-r}(x_{2},{\bf x})
d_{h'}^{*}(x_{3}, {\bf x}) |0\rangle \right\}\ , 
\label{link}
\end{eqnarray}
where ${\rm Vol}$ is the volume of transverse space; see figure 1.
The $M_{-+}$
boost-invariant  momentum fractions $x_1=k^{+}_{1}/P^+$ etc. have been
introduced, which should
not be confused with transverse co-ordinates.
DLCQ amounts
to discretizing such fractions into integer or odd half-integer units of $1/K$.
The wavefunction $|\psi(P^+)\rangle$ automatically satisfies the $k^+$
momentum and
helicity sum rules, even at finite $K$.

States of non-zero  ${\bf P}$ are straightforwardly generated by
application of $M_{-r}$
\begin{eqnarray}
\psi_{hh'}(x_1,x_2) 
& \to &  {\rm exp}\left[{\rm i} {\bf P}.{\bf x}\right]
\psi_{hh'}(x_1,x_2) \ , \nonumber \\
 \psi_{h(r)h'}(x_1,x_2,x_3) & \to &  {\rm exp}\left[{\rm i} 
{\bf P}.(x_1{\bf x} + x_2({\bf x} + 0.5a\hat{\bf r})
+x_3({\bf x}+ a\hat{\bf r}))\right] \psi_{h(r)h'}(x_1,x_2,x_3)  \, \\
\psi_{h(-r)h'}(x_1,x_2,x_3)& \to & {\rm exp}\left[{\rm i} 
{\bf P}.(x_1({\bf x}+ a\hat{\bf r}) + x_2({\bf x}+ 0.5a\hat{\bf r})
+x_3({\bf x}))\right]\psi_{h(-r)h'}(x_1,x_2,x_3) \ . \nonumber
\end{eqnarray}
A fixed Tamm-Dancoff approximation  will obviously lead to
Fock-sector-dependent renormalisations; in the present
case this means that $\mu_f$ is sector-dependent.
The analysis of later sections was performed both with and without
the assumption of sector dependence. The main conclusions are 
qualitatively similar, but only  results with a
sector-independent mass $\mu_f$ are shown, since it was possible to sweep the
coupling space with greater resolution in this case.

In the Fock space (\ref{link}),
the $x^-$ global colour-singlet
states have already been picked out
by applying the $J^+$ constraint (\ref{phys}), since it simplifies
the DLCQ Fock space. (Application
of the $F$ condition does not simplify the DLCQ Fock space
particularly, so is not imposed from the beginning; the
constraint will be satisfied dynamically when 
$P^-$ is diagonalised).
In the one-link truncation, a quark and antiquark either share
the same transverse lattice site, or are separated by one
link.  It is the simplest
approximation that still allows a meson to propagate on the
transverse lattice. 
A correct description of the
flavour-singlet sector would require inclusion of
 quark---anti-quark annihilation
processes, which are forbidden in the one-link approximation.
 Therefore, the  space  
(\ref{link}) will be interpreted as appropriate to the 
flavour-non-singlet mesons, by  identifying  
the quark and anti-quark with (degenerate) 
up and down flavours respectively. No
flavour label
is necessary in this case, and flavour is assumed to have been 
summed over in operators.

Projecting $2P^+ P^- |\psi ({\bf P}
= {\bf 0})\rangle$
onto Fock basis states, one derives the following set of coupled integral
equations for individual Fock components:
\begin{eqnarray}
{{\cal M}^2 \over \overline{G}^2} \psi_{hh'}(x_1,x_2) & = & 
\left( {m_{f}^{2} \over x_1}+ {m_{f}^{2} \over x_2} \right)
\psi_{hh'}(x_1,x_2) + K(\psi_{hh'}(x_1,x_2))
\nonumber \\
&& + {(k_{a}^{2} + k_{s}^{2}) \over \pi} \left( {1 \over x_1}
\int_{0}^{x_1} {dy \over y} +  {1 \over x_2} \int_{0}^{x_2} {dy \over y}
\right) \psi_{hh'}(x_1,x_2)              \nonumber \\
& & - \sum_{\lambda} \left\{ {m_f k_s \over 2 \sqrt{ \pi}} \int_{0}^{x_1}
        \frac{dy}{\sqrt{y}} \left(\frac{1}{x_1-y} + \frac{1}{x_1}
\right) \psi_{h(\lambda)h'}(x_1-y,y,x_2) \right. 
 \label{int1}    \\
& & \left. + {m_f k_s \over 2 \sqrt{  \pi}} \int_{0}^{x_2}
        \frac{dy}{\sqrt{y}} \left(\frac{1}{x_2-y} + \frac{1}{x_2}
\right) \psi_{h(\lambda)h'}(x_1,y,x_2-y) \right.
             \nonumber \\
& & + \left.  {{\rm Sgn}(\lambda) m_f k_a(h {\rm i} 
\delta_{|\lambda |1} + \delta_{|\lambda | 2}) \over 2 \sqrt{ \pi}} 
\int_{0}^{x_1}
        \frac{dy}{\sqrt{y}} \left(\frac{1}{x_1-y} - \frac{1}{x_1}
\right) \psi_{-h(\lambda)h'}(x_1-y,y,x_2) \right. 
  \nonumber    \\
& & \left. - {{\rm Sgn}(\lambda) m_f k_a( h'{\rm i} 
\delta_{|\lambda |1} +   \delta_{|\lambda |2})
  \over 2 \sqrt{ \pi}} \int_{0}^{x_2}
        \frac{dy}{\sqrt{y}} \left(\frac{1}{x_2-y} - \frac{1}{x_2}
\right) \psi_{h(\lambda)-h'}(x_1,y,x_2-y) \right\}     \nonumber \ ,
\end{eqnarray}


\begin{eqnarray}
{{\cal M}^2 \over \overline{G}^2} \psi_{h(\lambda)h'}(x_1,x_2,x_3) & = & \left(
{m_{b}^{2} \over x_2}  
+  {m_{f}^{2} \over x_1}+ {m_{f}^{2} \over x_3} \right)
\psi_{h(\lambda)h'}(x_1,x_2,x_3) + K(\psi_{h(\lambda)h'}(x_1,x_2,x_3))
\nonumber \\
& & - {m_f k_s  \over 2 \sqrt{  \pi x_2}} \left(
        \frac{1}{x_1} + \frac{1}{x_1 + x_2} \right) \psi_{hh'}(x_1+x_2,x_3) 
\nonumber \\
& & - {m_f k_s  \over 2 \sqrt{  \pi x_2}} \left(
        \frac{1}{x_3} + \frac{1}{x_2+x_3} \right) \psi_{hh'}(x_1,x_2+ x_3) 
\label{int2} \\
& &  - {\rm Sgn}(\lambda) {(h{\rm i}\delta_{|\lambda |1}+ 
\delta_{|\lambda |2}) m_f k_a  
\over 2\sqrt{  \pi x_2}} \left(
        \frac{1}{x_1} - \frac{1}{x_1 + x_2} \right) \psi_{-hh'}(x_1+x_2,x_3) 
\nonumber \\
& & + {\rm Sgn}(\lambda){(h'{\rm i}\delta_{|\lambda |1} + 
\delta_{|\lambda |2}) 
m_f k_a  \over 2 \sqrt{  \pi x_2}} \left(
        \frac{1}{x_3} - \frac{1}{x_2+x_3} \right) \psi_{h-h'}(x_1,x_2+
x_3)
\ . \nonumber 
\end{eqnarray}

The conventions
of ref.\cite{fran} have been adopted for the instantaneous gluon kernels
\be
K(\psi_{hh'}(x_1,x_2)) =  {1 \over 2 \pi} \int_{0}^{1} dy
\left\{ {\psi_{hh'}(x_1,x_2) - \psi_{hh'}(y,1-y) \over (y-x_1)^2}
	\right\} \ ,
\eq
\begin{eqnarray}
K(\psi_{h(\lambda)h'}(x_1,x_2,x_3)) & = & {1 \over 2 \pi} 
\int_{0}^{x_2+ x_3} dy  {(x_3 +2x_2 -y) 
\over 2(x_3 -y)^2\sqrt{x_2(x_2 + x_3 -y)}}\left\{
\psi_{h(\lambda)h'}(x_1,x_2,x_3) \right. \nonumber \\
&& \left.  - \psi_{h(\lambda)h'}(x_1,x_2 +x_3-y,y)  \right\}   
\nonumber \\
&&
+ {1\over 2 \pi x_3} \left(  \sqrt{1 + {x_3 \over x_2}} -1 \right)
\psi_{h(\lambda)h'}(x_1,x_2,x_3)
  \nonumber \\
&& + {1 \over 2 \pi} \int_{0}^{x_1+x_2} dy {(x_1 +2 x_2 -y)
\over
2(x_1-y)^2 \sqrt{x_2 (x_1 + x_2 -y)}} 
\left\{ \psi_{h(\lambda)h'}(x_1,x_2,x_3)
\right.  \nonumber \\ &&
\left.  -  \psi_{h(\lambda)h'}(y,x_1 + x_2-y,x_3) \right\}
\nonumber \\
&& + {1\over 2 \pi x_1 }\left( \sqrt{1 + {x_1\over x_2} } -1 \right) 
\psi_{h(\lambda)h'}(x_1,x_2,x_3)  \ . 
\end{eqnarray}
$\bar{G}=G\sqrt{(N^2-1)/N}$, which has the dimensions of
mass, and the following renormalisations have been introduced: 
\be
{\mu_b \over \bar{G}} \to m_b \ \ ; \ \ {\mu_{f} \over \bar{G}} \to
m_f \ \ ;   {\kappa_a \sqrt{N} \over \bar{G}} \to k_a \ \ ; \ \
{\kappa_s \sqrt{N} \over \bar{G}} \to k_s \ . \label{coup} 
\eq
The 
dimensionless variables $\{m_b, m_f, m_f k_s, m_f k_a \}$ appearing
in the Hamiltonian equations (\ref{int1})(\ref{int2}) are to be understood
as the renormalised couplings after imposing DLCQ and Tamm-Dancoff
cut-offs on the Fock space. All $N$-dependence of the theory has now
been absorbed into the couplings.

\subsection{Chiral symmetry}

Chiral 
symmetries are subtle both on a lattice and in light-cone
quantisation.
Consider first the effects of the lattice cut-off.
The lattice Lagrangian (\ref{ferlag}) explicitly  breaks
the chiral symmetry 
\be
\Psi \to e^{-{\rm i}  \theta \gamma_5}\Psi \ , \label{trans}
\eq
through the bare mass-term  $\mu_f$ and Wilson term $\kappa_s$.
The latter is needed to eliminate fermion doublers in the
free-fermion dispersion.

When chiral symmetry is not explicitly broken at the Lagrangian level
(a chiral theory),
spontaneous breaking would be signalled by
vacuum degrees of freedom in a conventional quantisation scheme. 
The light-cone problem with small
$k^+$ regulated does not have such explicit degrees of freedom in the
light-cone vacuum.
There are two, somewhat related, mechanisms one could envisage for 
the manifestation of spontaneously broken chiral symmetry in this case.
Zero modes, being vacuum degrees
of freedom,  may carry information
about the spontaneous breaking of chiral symmetry, and their
omission may require one to add counter-terms to $P^-$ that explicitly
break chiral symmetry, even if it is only broken
spontaneously in the conventional sense \cite{wilson1}.
Spontaneous breaking may occur in a more-or-less
conventional way through wee partons 
that enhance contributions from operators that break chiral
symmetry in the limit that their couplings  are sent to zero \cite{suss}.
An explicit illustration of the latter effect was given in
ref.\cite{mat3}. 
Note that,  once a Tamm-Dancoff cut-off is introduced, the effects of
both these mechanisms on the Hamiltonian would be similar. The wee
partons
are excluded by this cut-off, so the net result of the second 
mechanism would be to introduce finite 
couplings to chiral-symmetry breaking counterterms in the cut-off theory.

At the
Lagrangian level, the chiral limit is $\mu_f \to 0$ and $\kappa_s \to
0$. However, the renormalised
parameters $\{m_{f}^{2}, m_f k_s, m_f k_a \}$
 need not be zero in the chiral limit.
To address chiral symmetry at this level, one can 
introduce the light-cone chiral transformation \cite{wilson1}  
\be
\Psi^+ \to e^{-{\rm i} \theta \gamma_5}\Psi^+ \ , \label{hel}
\eq
where $\Psi = (\Psi^+ + \Psi^-)/\sqrt{2}$ has been defined earlier,
and $\Psi^-$ is determined through the constraint equation (\ref{cons}).
This is equivalent to the usual chiral transformation on $\Psi$
(\ref{trans}) only
in a chiral theory.
Omitting zero modes, the light-cone chiral charge that generates
(\ref{hel}) measures helicity
\begin{eqnarray}
Q_{5} & = & \int dx^- \sum_{\bf x} \overline{\Psi} \gamma^{+} \gamma_5 \Psi
\nonumber \\
& = & \int dx^- \sum_{\bf x} \sum_{h} h u^{*}_{h} u_{h} \ . 
\end{eqnarray}
It is equivalent to the usual chiral charge only in a chiral theory
without spontaneous breaking, the difference presumably lying in
contributions at $x^-$ infinity. In a chiral theory with spontaneous
breaking of the symmetry and no explicit small $k^+$ degrees
of freedom to reflect this,  one therefore expects 
helicity-violating operators in the renormalised Hamiltonian.
In the Hamiltonian equations (\ref{int1})(\ref{int2}), 
only the operator coupling to 
$m_f k_a$ breaks the helicity symmetry (\ref{hel}).

Conservation of the renormalised axial current $A_{\mu}$ defines a chiral
theory. One may use PCAC to impose chiral symmetry on the theory,
through
the matrix element
\be
\langle 0| \partial_{\mu} A_{\mu} |\psi_{\pi}(P^{\mu})\rangle  =  f_{\pi}
{\cal M}_{\pi}^{2} \ .
\eq
Without knowing the precise expression for $A_{\mu}$ in terms of bare
fields, chiral symmetry can be imposed by setting either
$f_{\pi} = 0$ or ${\cal M}_{\pi} = 0$. To test for
a phase of gauge theory with spontaneously broken chiral symmetry, one would
then search the space of couplings for a 
region where Poincar\'e symmetry is restored, ${\cal M}_{\pi} = 0$,
and there is no instability. 
In the leading approximation considered in this paper, $m_f k_a$
couples to the only hopping term that propagates
the $\pi$ and $\rho$ at different rates on the transverse lattice.
In the absence of any other hopping term that distinguishes the
pion and rho, Poincar\'e 
symmetries would require this coupling to be as small as possible,
i.e. as small as stability of the theory allows.
$m_f k_a$ is also the only
source of mass difference between the pion and rho.
Unfortunately,  stability is difficult to check from first
principles in the one-link approximation. Any instability
in the quantum field theory would almost certainly be signalled
by copious parton production, which is forbidden in the one-link
approximation.

Because of the difficulty in checking for instability in the
approximation made in this paper,
a phenomenological constraint will be introduced
that is intended to model the effects of spontaneously broken
chiral symmetry in the one link approximation.
First, the experimental 
string tension $\sigma$ will be used to set the QCD scale.
This quantity was calculated in terms of lattice quantities in 
ref.\cite{dv2}.\footnote{These 
calculations involved no modelling, using 
Poincar\'e symmetry and the power expansion 
alone. The large-$N$ limit was also taken. However, the
present calculation is independent of $N$ in the one-link approximation
up to a trivial redefinition of $G$. 
$G\sqrt{N}$ in ref.\cite{dv2} is identified with $\bar{G}$.}
Then both the 
$\pi$ and  $\rho$ masses will be fit to the experimental values.
This involves three fits to experimental numbers, when  
QCD itself has only two parameters (e.g. the quark mass and
$\Lambda_{QCD}$).
Thus there is a single phenomenological constraint.
The use of $\pi$ and  $\rho$ masses 
is intended to model
the fact that (a) a nearly chiral theory in the Goldstone mode 
should  have a light pion
and (b) helicity violation, that splits the $\pi$ and $\rho$ masses,
should be present in $P^-$ when chiral symmetry is broken only
spontaneously.
The remaining freedom in the couplings present in $P^-$ will be
fixed by optimizing meson dispersion. This, together with the use of a
more accurate calculation of $\sigma$, represents
the main advance over the purely phenomenological modelling 
of this lattice gauge theory performed in ref.\cite{mat0}.

\section{Model Calculations}

\subsection{Symmetry Tests}

The one-link approximation is too simple to allow much 
change in a meson's  size in lattice units, while retaining its physical
size and covariant dispersion. Therefore,  the spacing $a$ will be
fixed
at the
largest value that gave reasonable results in ref.\cite{dv2},
by setting $m_b =0.2$. In this case $\bar{G} \approx 3.11
\sqrt{\sigma}$
and $a \bar{G} \approx 4.57$. If one takes $\sqrt{\sigma} = 440$ MeV, then 
$\bar{G} = 1368$ MeV and $a \approx (300 {\rm MeV})^{-1} \equiv 2/3$ fm.
This is obviously the maximum quark separation in the one-link  
approximation.

Since the Hamiltonian (\ref{lcham}) is transversely 
local, it makes sense to expand eigenvalues thus
\be
2P^+ P^- = {\overline G}^2 \left( {\cal M}^{2}_{0} 
+ {\cal M}_{1}^{2}\, a^2
     {\bf P}^2 + O(a^4 {\bf P}^4)  \right) \label{latshell}\; .
\eq
A necessary condition for Poincar\'e invariance is
isotropy of the dispersion, i.e. the  speed of light $c$ in transverse
directions should satisfy
\be
a^2 \overline{G}^2 {\cal M}_{1}^{2}  \equiv c^2  =  1 \label{cond}  \ .
\eq
It makes no sense to try to tune the anharmonic terms in
(\ref{latshell}) separately, 
since only nearest neighbour hopping is available in the current 
approximation.

The transverse lattice has symmetry group $D_4$ \cite{bard2}, with 
one-dimensional representations and a single two-dimensional
irreducible representation.
Since there is $90^{\rm o}$ rotational symmetry about $x^3$, it is 
possible to distinguish the angular momentum projections
$J_3$ mod 4. The one-dimensional representations corresponds to $ J_3
= 0$ or symmetric and antisymmetric combinations of $ J_3= \pm 2$.
The two-dimensional
contains $J_3 = \pm 1$. Generically, the pion is 
split from the rho, while the $J_3 = 0$ component of the rho
($\rho_0$) is
split from its degenerate $J_3 = \pm 1$ components ($\rho_{\pm}$). However, all
these states become degenerate in the absence of quark helcity
violating interactions.
The zero-link sector serves to distinguish the various
meson components, since they do not mix when ${\bf P}={\bf 0}$ as a
result of $D_4$ symmetry.
\begin{eqnarray}
& \pi : & \psi_{+-} = - \psi_{-+} \ ; \ \psi_{\pm \pm} = 0
\ . \nonumber \\
&\rho_0 : &\psi_{+-} = \psi_{-+} \ ; \ \psi_{\pm \pm} = 0 \ . \\
&\rho_{\pm}: &\psi_{\pm \pm} \neq 0 \ ; \ \psi_{+-} =
\psi_{-+} = 0\  .
\end{eqnarray}

The three-dimensional space of dimensionless 
couplings $\{m_f, k_a, k_s \}$ was sampled discretely and the 
eigenvalue problem for $2P^+ P^-$ solved for various ${\bf P}$.
A $\chi^2$-test was introduced for the conditions (\ref{cond}) on the
$\pi$ and $\rho$ dispersion 
and for the conditions that their masses assume the physical
values,  ${\cal M}_{\pi}  = 140$ MeV and ${\cal M}_{\rho} = 770$ MeV;
\be
\chi^2 =  w_1 \mid {{\cal M}_{\pi} \over {\cal M}_{\rho}} - 0.18 \mid + 
w_2 \mid{{\cal M}_{\rho} \over \overline{G}} - 0.56\mid
+  \sum_{\pi, \rho_{\pm},
\rho_{0}}  |c-1| \ .
\eq
Masses are defined as $\overline{G} {\cal M}_{0}$ from (\ref{latshell}), where
${\cal M}_{\rho}$ is taken to be the 
average of the $\rho_0$ and $\rho_{\pm}$ masses.
The weights $w_1$ and $w_2$  were adjusted to allow tolerance
of about $100$ MeV in satisfying the mass conditions. 
The following results used $w_1 = w_2 = 10$.
Calculations were performed for DLCQ cut-offs $K=6,7,8,9$, with
the corresponding bases of dimension $264,364,480,612$. The size of Fock 
space is quite modest and there is ample room for larger and more
efficient calculations to be performed. 

Figure 2 shows the results of the $\chi^2$ test. A clear minimum
appears for every coupling at every $K$. While the details will change
slightly with the specific test used, the broad features remain the
same. To indicate what can be achieved at the minima of the $\chi^2$ shown in
fig. 2, at $K=9$ for example,  the observables are shown in Table
1.
For later calculations, it is convenient to fit the mid-point of the 
$\chi^2$ valley at each $K$  to 
a smooth form $A + B/K$,
\begin{eqnarray}
m_f & = & 0.362(108) - {0.969 \over K}  \ ,\nonumber \\
k_a & = & 0.162(927) + {8.34 \over K} \ , \label{smooth} \\
k_s & = & -0.323(921) + {8.29 \over K}  \ . \nonumber
\end{eqnarray}
This will represent the  best estimate of the symmetry-restoring trajectory
at each $K$, for the range $K=6 \to 9$.
When fitting to polynomials in $1/K$,  the magnitude of the
last term in the series will be taken as an estimate of the error on the
$K \to \infty$ result (the zeroth order  term of the polynomial).
Obviously,  the fit to the couplings (\ref{smooth}) does not produce a very
useful estimate of the their $K \to \infty$ limit. However,
the forms (\ref{smooth}) are only used in the range $K=6 \to 9$ to recompute
observables at those values, with a view to extrapolation of the
observables themselves, whose convergence is typically  faster
than that of the couplings.

The observables were also fit to linear or quadratic polynomials in  $1/K$. 
A series in $1/K$ is the 
correct form for the errors introduced by DLCQ discretization
of the integral equations (\ref{int1})(\ref{int2}), 
if the integrands are analytic
and the couplings are polynomial in $1/K$.
The integrands are in general not analytic, which  will also produce 
non-integral powers of $1/K$
in the finite-$K$ errors. However, 
the values of $K$ used here are too few and too small to warrant
this level of complication in fitting functions.
\footnote{The errors introduced by extrapolation in $K$ may make 
DLCQ seem inefficient compared to a basis of smooth wavefunctions. 
However, DLCQ is superficially 
 easier to extend to multi-parton Fock states, which must be the 
eventual goal.}

The light-cone eigenfunctions  were recomputed along the
smooth trajectory (\ref{smooth}) and used to derive a number of
observables for the $\pi$ and $\rho$. The
observables
are usually quoted at some transverse scale, which in the case
of the tranvserse lattice  is provided by $Q = \pi / a \sim 1$ GeV,
the normalisation scale at which the non-perturbative
wavefunctions have been determined.
Observables will in general evolve with this scale, but $a$
has been fixed for the present calculation.
The relationship between this scale and those introduced in other
quantisation schemes ($\overline{\rm MS}$ etc.) is non-trivial
and will not be addressed here. For estimates, 
$Q$ will be identified with the momentum transfer  from the
probe where appropriate.

\subsection{Distribution amplitudes}
 
Distribution amplitudes for the $\pi$  and $\rho$ are defined (with
covariant
normalisation) by
\begin{eqnarray}
\langle 0| \overline{\Psi}(z) \gamma^{\mu} \gamma_5 \Psi(0)
|\psi_{\pi}(P^{\mu})\rangle |_{z^2 = 0} & = & f_{\pi} P^{\mu} \int_{0}^{1}
{\rm e}^{{\rm i} x (z.P)} \phi_{\pi}(x) \label{pidist} \ , \\
\langle 0| \overline{\Psi}(z) \gamma^{\mu}  \Psi(0)
|\psi_{\rho_{0}}(P^{\mu})\rangle |_{z^2 = 0} & = & f_{\rho_0} P^{\mu} \int_{0}^{1}
{\rm e}^{{\rm i} x (z.P)} \phi_{\rho_0}(x)  \ , \\
\langle 0| \overline{\Psi}(z) \sigma_{\mu \nu}  \Psi(0)
|\psi_{\rho_{\pm}}(P^{\mu})\rangle |_{z^2 = 0} & = &  
f_{\rho_{\pm}} (\epsilon^{\pm}_{\mu}(P) P_{\nu} -
\epsilon^{\pm}_{\nu}(P) P_{\mu}) \int_{0}^{1}
{\rm e}^{{\rm i} x (z.P)} \phi_{\rho_{\pm}}(x) \ ,
\end{eqnarray}
where the $\phi$'s are normalised to one and $\epsilon$ is the
polarization vector.
They govern the leading-order
perturbative QCD expression for many high momentum transfer exclusive processes
\cite{stan1}. 
Current operators appearing in matrix elements, such as those 
above, will require further renormalisation 
to make them  covariant in the presence of
cut-offs, in the same fashion as the Poincare generators.
In this paper, consistent with similar approximations made on the
hamiltonian and Hilbert space,
the bare current operators valid to lowest order of the power
expansion in dynamical fields will be used to estimate matrix elements.

For the pion, one finds
\be
\psi_{+-}(x,1-x) =   {f_{\pi} \over 2} \sqrt{\pi \over N}  \phi_{\pi}(x)  
\eq
from the $\gamma^+$ component of (\ref{pidist}). 
Figure 3 shows this distribution amplitude  at various
 $K$.  A cubic
interpolating function is used to produce a curve from the
discrete values of momentum fraction 
$x$ sampled in DLCQ at each $K$. The resulting
curves are extrapolated pointwise with a fit to a quadratic in
$1/K$, to yield the data points with extrapolation error bars shown
also in figure 3. These extrapolated data points have been fit to the
first two terms of the conformal expansion \cite{conf1,conf2}
\be
\phi_{\pi}(x)  =  6 
x (1-x) \left\{ 1 + 0.133 C_{2}^{3/2}(1-2x^2)\right\} 
 \ . \label{exppi}
\eq
The overall normalisation yields $f_{\pi} = 347(46)$ MeV,
compared to the experimental value $f_{\pi}({\rm exp.}) = 131$ MeV. 
Because of error in the MeV scale setting, in the extrapolation in $K$, 
in the proportion of valence quarks
in the pion, and/or in renormalisation,
the result is much higher  in the
current approximation, compared to the real world. 
One would expect that there are too many valence quarks at zero
transverse separation in the one-link approximation, overestimating
electroweak decay constants.

Direct experimental measurements of $\phi_{\pi}(x)$ 
have recently become possible
from diffractive dissociation on a nucleus $\pi + A \to A + {\rm
jets}$ \cite{ashery}. These experiments are carried out at 
somewhat larger $Q^2 \sim 10 {\rm GeV}^2$, to minimize initial
state interactions. Although the 
normalisation, which is $Q^2$-independent, is not correct,
leading-order perturbative QCD evolution with boundary condition
(\ref{exppi}) produces a curve  with shape consistent with the data measured
in ref.~\cite{ashery}.
The moments of $\phi_{\pi}$ have also been estimated from
QCD sum rules. The most recent results \cite{nlcsr} indicate a
distribution amplitude close to the asymptotic form
$x(1-x)$, even at low scales; this result is completely different
from the early sum rule calculations of Chernyak and Zhitnistky
\cite{cz}, who suggested a large positive $C_{2}^{3/2}$ coefficient.
The second moment of $\phi_{\pi}$ has also been calculated
from first principles in Euclidean lattice QCD
\cite{lattice}. Although there is some systematic variation in
those results, assuming the form (\ref{exppi}),
the most recent lattice results
suggest a somewhat negative $C_{2}^{3/2}$ coefficient! 
This also appears consistent with the  E791 data, provided the
coefficient is not too large.

An analysis for the $\rho$ yields very similar shapes 
and normalisations for the distributions. 
For the helicity zero component of the rho, for example, one finds
 $f_{\rho_{0}} =315(50)$ MeV, again much more than
the experimental value $f_{\rho_{0}}({\rm
exp.})= 216$ MeV.

\subsection{Parton distributions}

The higher Fock state structure of the light-cone wavefunctions in 
general plays a role in any low momentum transfer process as well as
high momentum transfer inclusive processes.
For example, deep inelastic scattering probes
the parton distribution function, the probability for finding a
quark
carrying certain longitudinal momentum fraction of the hadron:
\be
V(x,Q^2)  =   \sum_{h, h'} |\psi_{hh'}(x,1-x)|^2 
 + \sum_{\lambda}\sum_{h,h'}
 \int_{0}^{1-x} dy  \  |\psi_{h(\lambda)h'}(x,y,1-x-y)|^2  \ .
\eq
The quark distribution function calculated in the
pion is shown in figure 4. In this case a linear extrapolation in $1/K$
was necessary to achieve a smooth distribution. 
Also shown is a relevant phenomenological distribution that neglects
sea quarks \cite{grv}, 
\be
V(x,0.23) = {0.526 \over x^{0.495}}(1+ 0.357 \sqrt{x})(1-x)^{0.365} \ ,
\eq 
which was produced by  fitting  experimental $\pi N$ scattering data, after 
perturbative evolution to higher scales.
(Note there is a slight difference in 
$Q^2$ scale).
Although the calculated parton  distribution appears close to the 
one of ref.\cite{grv}, the excess at large $x$ means that the
calculated total
quark momentum fraction (the first moment of $V(x)$), $0.43(1)$, is 
higher than typical experimental estimates \cite{na3}. 
Similarly, it is higher than 
estimates
from quenched Euclidean lattice calculations of the
low moments of $V$ \cite{mom}, which yield about 0.3 for the first
moment
(at a slightly higher scale). 
This is consistent with the fact that the neglect of higher numbers
of links in the one-link approximation
excludes many gluon channels,  so the quarks must pick up the
momentum instead. 

The pion elastic electromagnetic form factor depends upon 
a coherent overlap of the entire light-cone wavefunction at low
momentum transfer: 
\be
\langle\psi(P')|  \int dx^- {\em e}^{{\rm i} q^+ x^-}
\sum_{\bf x} {\em e}^{{\rm i} {\bf q}.{\bf x}}
\sum_{\eta} e_{\eta} \overline{ \Psi}_{\eta} \gamma^{\mu} \Psi_{\eta} 
|\psi(P) \rangle =  F(-q^2){(P'+P)^{\mu} \over 2}\ ,
\eq
where $q = P' - P$ is the exchanged photon's four-momentum and 
a flavour label $\eta$ and charges $e_{\eta}$ have been explicitly
introduced to avoid
confusion.
The space-like form factor $F(-q^2)$, $-q^2 > 0$, can be computed
most simply from the $+$ component of the current 
in the Drell-Yan frame $q^+ = 0$ \cite{drell}, in which 
the photon's momentum is purely transverse.
One finds
\begin{eqnarray}
F(-q^2) & = & \int_{0}^{1} dx_1 dx_2
\sum_{h,h'} \delta (x_1 + x_2 -1) |\psi_{hh'}(x_1,x_2)|^2  \\
&& + 2\sum_{h,h',r} \int_{0}^{1}dx_1 dx_2 dx_3 \delta (x_1 + x_2 +x_3 -1)
{\rm Cos}\left( (x_1 + {x_2 \over 2})a {\bf
q}.\hat{\bf r}
\right) 
|\psi_{h(r)h'}(x_1,x_2,x_3)|^2 \ . \nonumber 
\end{eqnarray}
$G$-parity and transverse reflection symmetries have
been used to simplify the result.
The resulting form factor, with ${\bf q}$ along a lattice axis,
is shown in figure 5 (the calculated $F$ suffers from lattice artifacts
 and so is not rotationally invariant). 
The charge radius, given by $r_{\pi}^{2} = 6 (\partial F(-q^2)/\partial
q^2)_{q^2 = 0}$, 
will obviously be artificially reduced
 compared to the experimental value, $r_{\pi} = 0.663$ fm \cite{rpi}.
This distance is roughly the same as one lattice spacing, which is the
{\em maximum} quark separation in the one-link approximation. This
explains why the curve in fig.5 is  flatter than the true one
\cite{bebek}; the slope at the origin gives a charge radius about
half the experimental value.

The
quark distribution functions in the $\rho$ are shown in figure 6.
One interesting feature, which may persist in more accurate
calculations, is a difference between the longitudinal and
transversely
polarized $\rho$. This is
consistent
with some of the $\rho$'s spin coming from orbital angular
momentum of the quarks. In fact, one can directly measure the quark
spin distribution in a transversely polarized rho meson $\rho_{\pm}$
(it is identically zero in a $\rho_0$).
Define the helicity asymmetry
\be
A(x) = {V_{+}(x) - V_{-}(x) \over V_{+}(x) + V_{-}(x)} \ ,
\eq
where $V_{\pm}(x)$ is the probability for a quark with helicty aligned
or anti-aligned with that of the polarized $\rho$ respectively;
$ V(x) = V_{+}(x) + V_{-}(x)$.
Figure 7 shows that a large part of the $\rho$ spin is not
carried by the quark spins; one finds $s=0.41(10)$ for the quark spin fraction,
\be
s= \int_{0}^{1} dx \ \left( V_{+}(x) - V_{-}(x) \right) \ .
\eq
In the non-relativistic quark model, $s=1$. About half the rho's spin 
therefore must come from orbital angular momentum and gluon
spin.
Furthermore, fig. 7 exhibits the phenomena of helicity
retention --- the quark tends to align its spin with that
of the hadron at large $x$ --- which is expected on general
grounds \cite{hr}. At small $x$ the quark helicity tends to become
disordered, and the asymmetry $A(x)$ may even change sign eventually
\cite{ladder}.
Euclidean lattice estimates of the low moments of 
the rho parton distributions \cite{mom} 
are consistent with the above results, except for the fact that, as
with the pion, too much momentum is carried by quarks in the
one-link approximation.

There are a number of subtleties in extracting the electromagnetic
form factors of
vector mesons from the light-cone wavefunctions; see for example,
the discussion in ref.\cite{melo}. Since realistic values will in any case
not be obtained in the one-link approximation,  an
analysis will not be attempted here for the $\rho$.

\section{Conclusions}

Knowledge of the light-cone wavefunctions provides evident advantages 
over other methods, which determine low moments of operator matrix
elements. Apart from renormalisation of the operators
themselves, which is common to all  methods,
no further significant effort is required to calculate
the matrix elements once the light-cone wavefunctions are 
determined. Moreover, these matrix elements are determined directly as
functions of the momenta involved. On the other hand, identical low moments
can give rise to functions of very different shape
and moment integrals  can often be very sensitive
to inaccessible corners of phase space.

This paper has begun a program  to calculate the light-cone wavefunctions of
conventional hadrons from a transverse lattice gauge theory, using
symmetry principles to renormalise the light-cone Hamiltonian.
Expanding the most general Hamiltonian and the Fock space
in powers of dynamical fields,
the leading approximation led to practical difficulties 
in imposing simultaneously both Poincar\'e and chiral symmetry 
from first principles. While Poincar\'e symmetry could be tested
through dispersion and PCAC could be imposed through ${\cal M}_{\pi}$,
tests for stability of the theory would be compromised by the
one-link approximation.
In the $\pi$-$\rho$ sector, the mass splitting was therefore used as
a  phenomenological constraint to model the
realisation of chiral symmetry in spontaneously broken form.
The condition of isotropic dispersion then proved sufficient
to unambiguously fix the remaining couplings appearing in the
leading approximation.
The resulting  calculation in a Fock space restricted to at most one link
 reproduced a number of 
features of the pion and rho, known from experiment or
estimated  from Euclidean lattice calculations.

The next steps, which must go beyond the simplest approximations
made here, will allow two questions to be addressed.
Can symmetry principles alone be used to unambiguously renormalise
the transverse lattice theory within a given approximation?
Does the approximation scheme converge rapidly-enough to be of
practical use?
By extending the Tamm-Dancoff cut-off beyond one-link, it becomes
easier to detect instabilities in the theory. As discussed in section~II.D,
this is one means by which chiral symmetry could be addressed from
first principles --- requiring a massless pion in a stable, Poincar\'e
invariant theory.
Beyond the one-link approximation it also becomes
meaningful to add further quartic operators to the Hamiltonian.
This will introduce further sources of parity and helicity
violation that may be used to restore some of the parity
and chiral symmetry broken by cut-offs. The Poincar\'e and chiral
symmetry restoration should then be demonstrated over a range of lattice
spacings $a$, as has been done for Poincar\'e symmetries in the
case of pure gauge theories.

From a phenomenological point of view, the one-link 
approximation, with a lattice spacing of 2/3 fm, artificially
squashes light mesons. By adding more links, 
allowing these mesons to expand to their physical size,
observables such as form factors and quark
momentum fractions should move closer to their experimental values.
It will also become reasonable to include $P$-wave mesons in the
analysis, with  more symmetry constraints coming from their
dispersion. Moreover, the simplest approximation for propagating baryons
on the transvere lattice will require configurations with two links.

\vspace{10mm}
\noindent {Acknowledgements}: I thank M. Burkardt for
useful
discussions. This work was supported by PPARC grant GR/LO3965.

\newpage

\begin{table}
\centering$\displaystyle
\renewcommand{\arraystretch}{1.25}
\begin{array}{|c|cccc|}
\hline
 K=9 & \pi  & \rho_{0} & \rho_{+}  & \rho_{-} \\
\hline
{\cal M} {\rm (MeV)}  & 153 & 848  & 601 & 601  \\
c & 1.18 & 0.73 & 0.69 & 1.16\\
\hline
\end{array}$
\caption{Observables at the minimum $\chi^2$ for $K=9$.
\label{table1}}
\end{table}

\newpage

\begin{center}
CAPTIONS
\end{center}

FIGURE 1 - Transverse lattice (solid lines) at fixed $x^+$ and
 meson configurations in a one-link
approximation. Chain lines represent $P \ {\rm exp} \int dx^- A_{-}$
required for gauge invariance.

FIGURE 2 - $\chi^2$ as a function of couplings for $m_b = 0.2$ 
and $K=6,7,8,9$.

FIGURE 3 - The distribution amplitude $\psi(x) = \psi_{+-} = -
\psi_{-+}$  of the pion, at a
transverse normalisation scale $Q^2 \sim 1$ $ {\rm GeV}^2$. Grey
curves correspond to DLCQ cut-offs $K=6,7,8,9$ 
(darker means larger $K$). Data points are the pointwise extrapolation
of finite-$K$ curves. The solid curve fits these data to the 
conformal expansion (\ref{exppi}).

FIGURE 4 - The quark distribution function in the pion at a
transverse normalisation scale $Q^2 \sim 1 {\rm GeV}^2$. 
Grey
curves correspond to DLCQ cut-offs $K=6,7,8,9$ 
(darker means larger $K$). Data points are the pointwise extrapolation
of finite-$K$ curves. The solid curve (GRV) comes from a  
phenomenological fit to $\pi N$ scattering data \cite{grv}, at
normalisation scale $0.23$ ${\rm GeV}^2$.

FIGURE 5 - The pion electromagnetic form factor $F(-q^2)$. Grey
curves correspond to DLCQ cut-offs $K=6,7,8,9$ 
(darker means larger $K$). Data points are the pointwise extrapolation
of finite-$K$ curves.

FIGURE 6 - The quark distribution function in the longitudinally
polarized $(h=0)$ and transversely polarized $(h=1)$
rho meson, linearly extrapolated in $1/K$.

FIGURE 7 - The quark helicity asymmetry $A(x)$ in a transversely polarized
$\rho$, linearly  extrapolated in $1/K$.

\begin{figure}
\centering
\BoxedEPSF{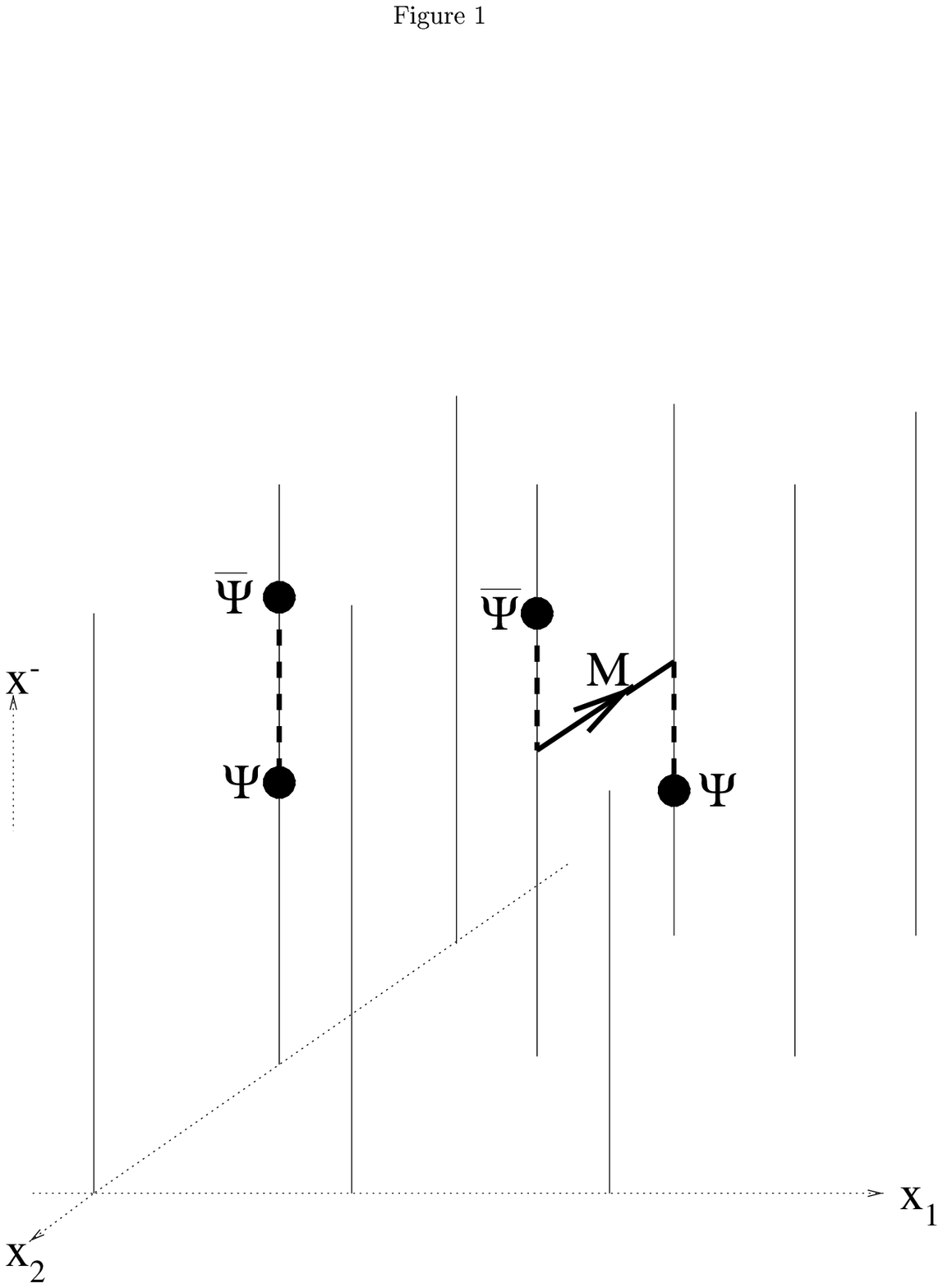 scaled 800}
\label{fig1}
\end{figure}

\begin{figure}
\centering
\BoxedEPSF{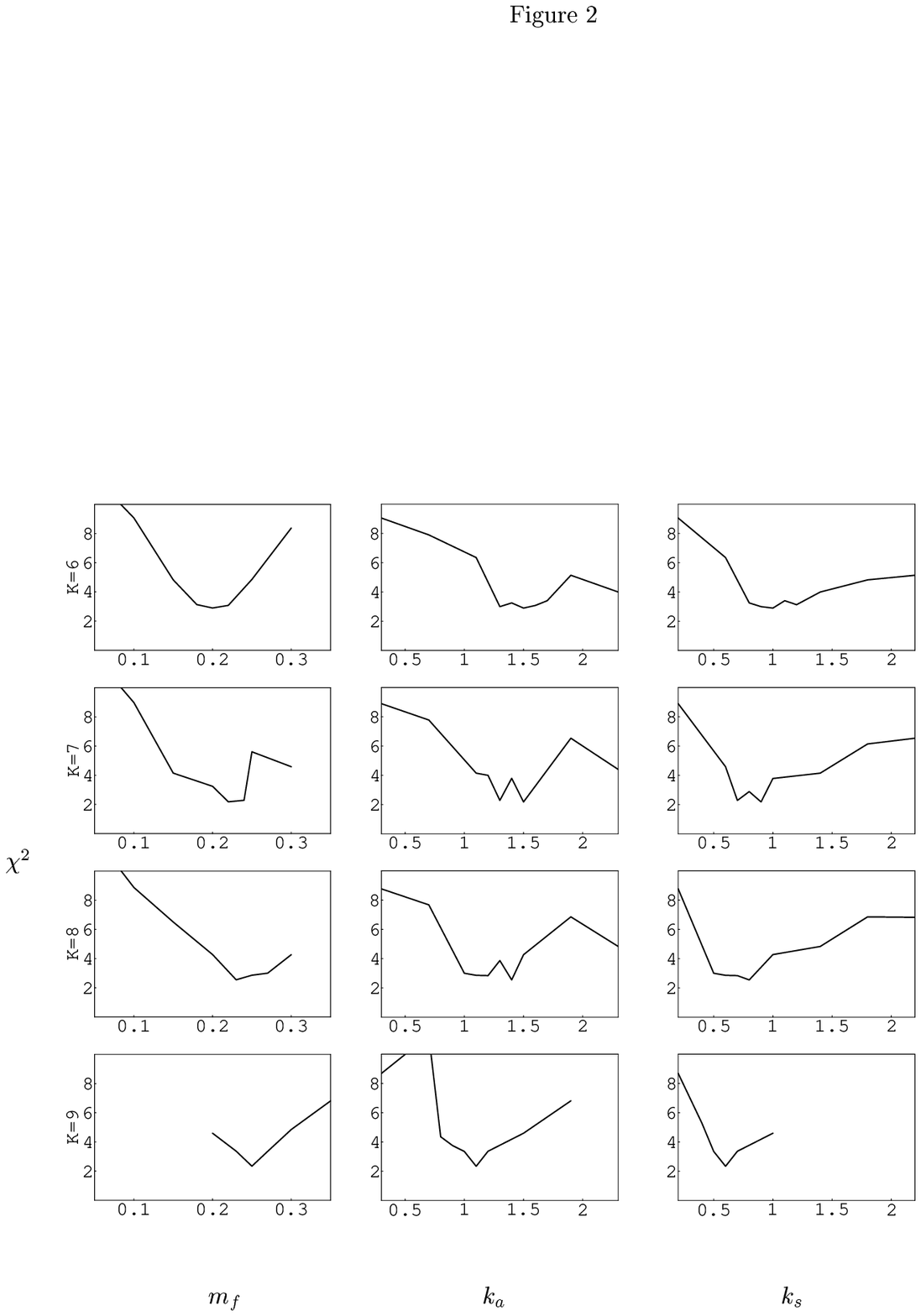 scaled 800}
\label{fig2}
\end{figure}

\begin{figure}
\centering
\BoxedEPSF{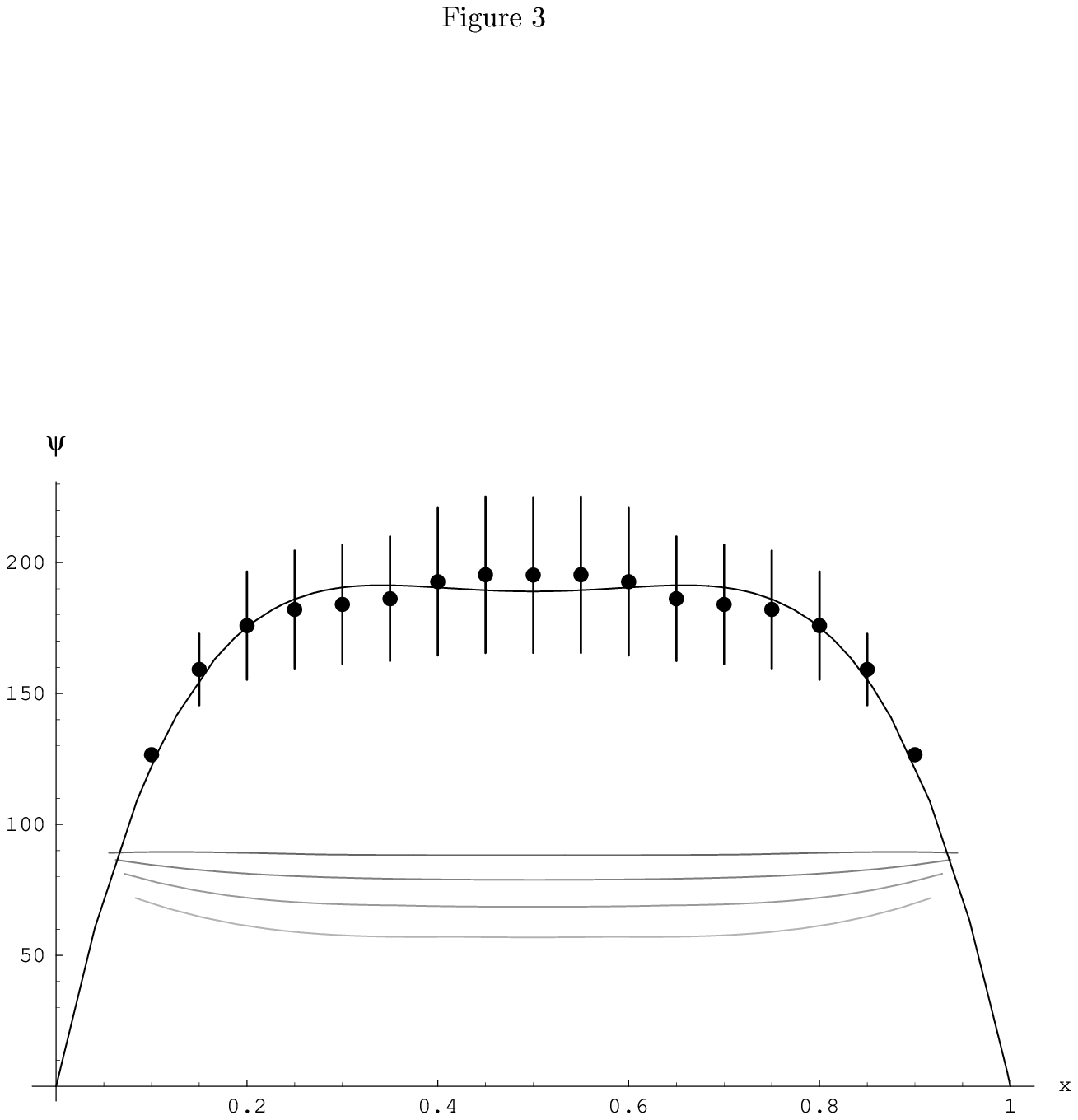 scaled 800}
\label{fig3}
\end{figure}

\begin{figure}
\centering
\BoxedEPSF{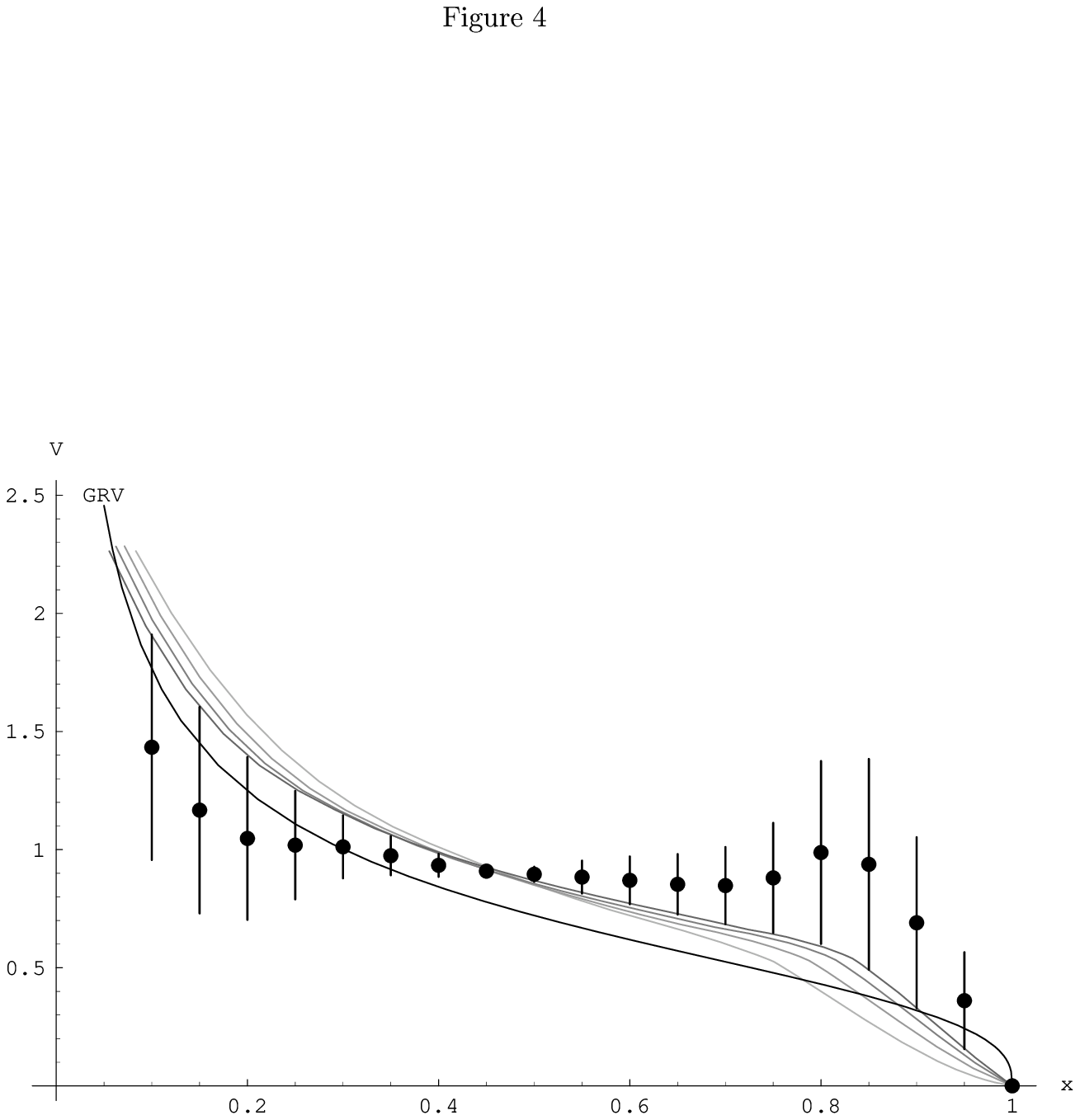 scaled 800}
\label{fig4}
\end{figure}

\begin{figure}
\centering
\BoxedEPSF{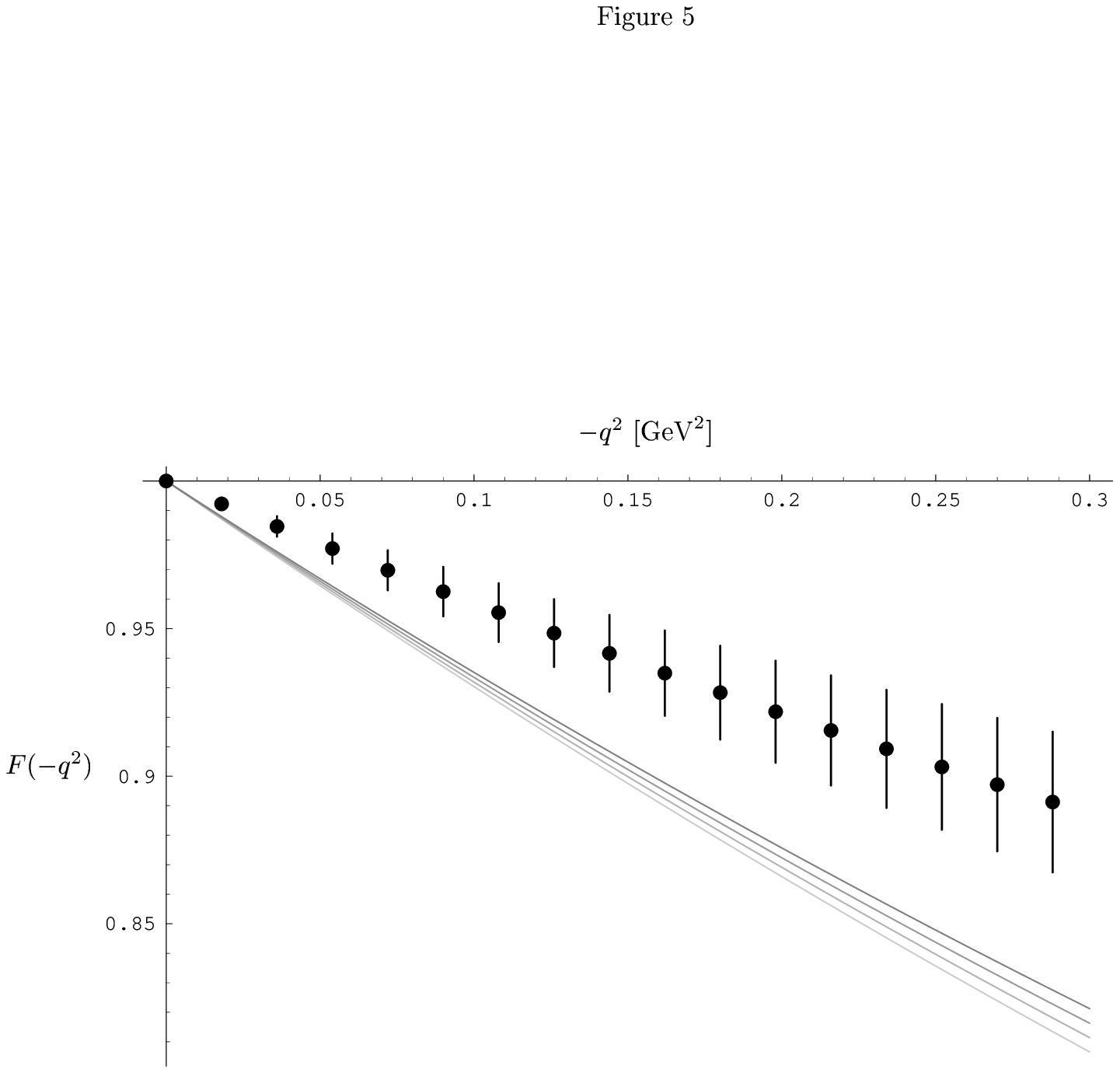 scaled 800}
\label{fig5}
\end{figure}

\begin{figure}
\centering
\BoxedEPSF{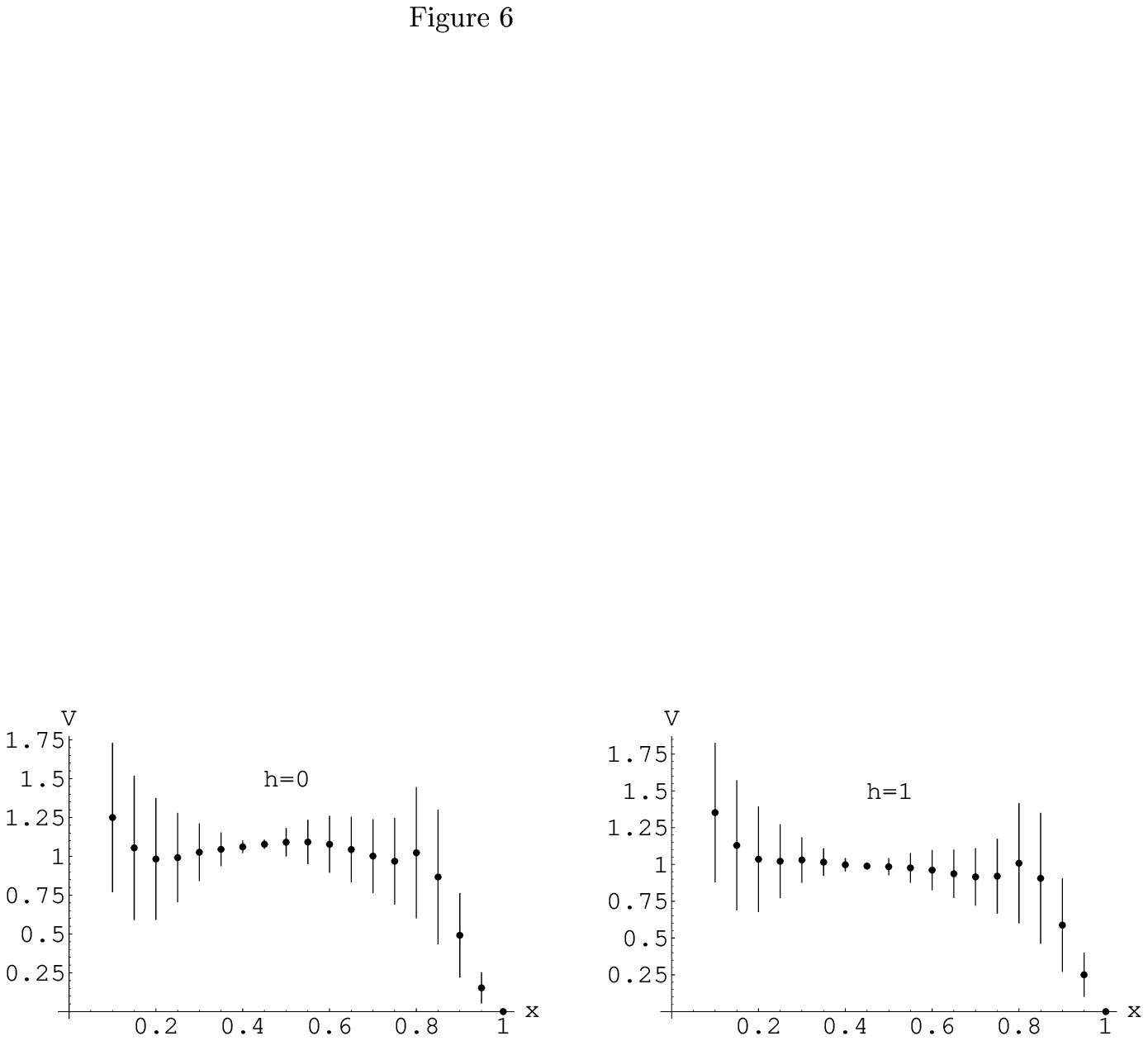 scaled 800}
\label{fig6}
\end{figure}

\begin{figure}
\centering
\BoxedEPSF{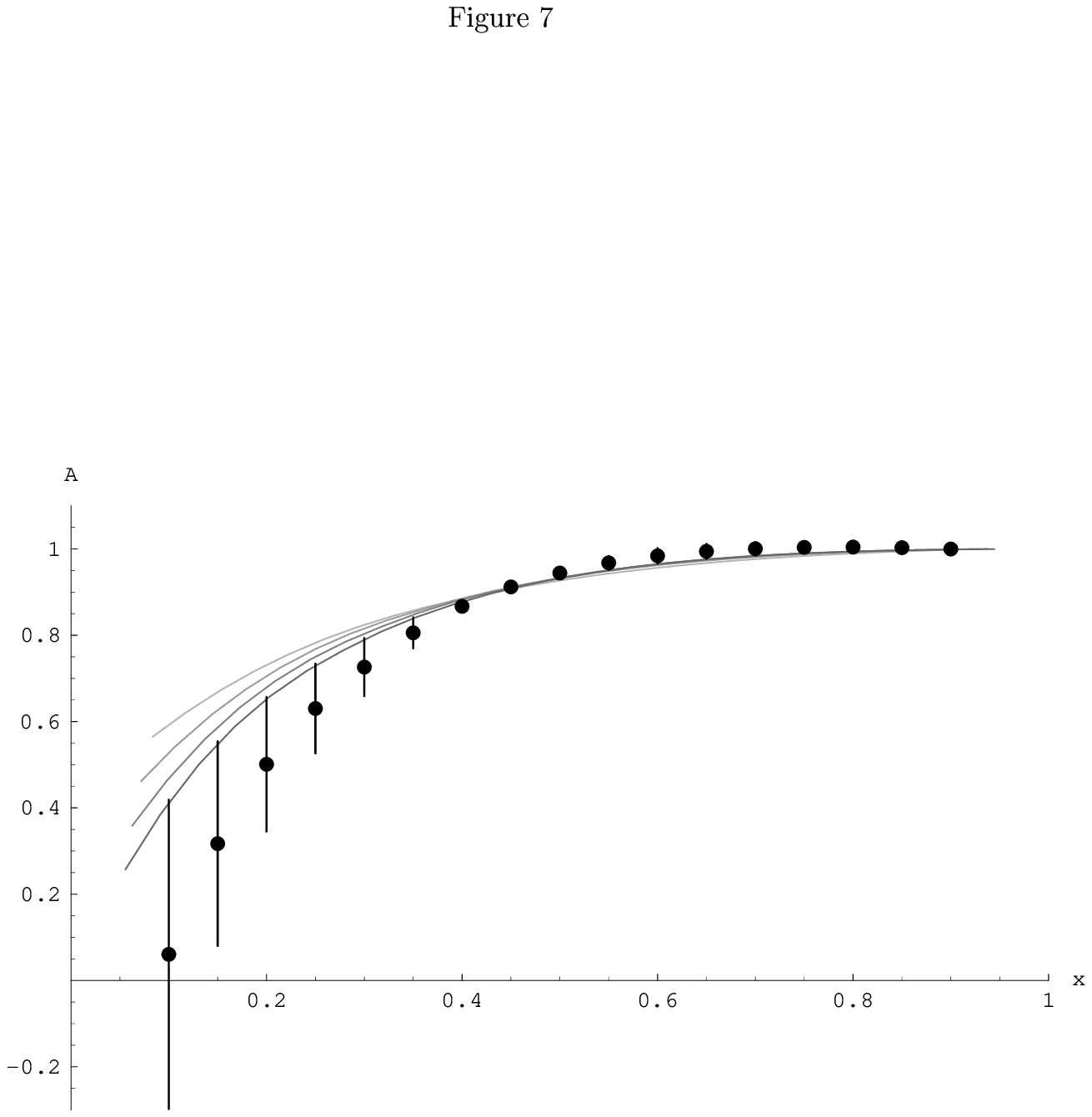 scaled 800}
\label{fig7}
\end{figure}

\end{document}